\documentclass[11pt, a4paper, fleqn]{article}

\usepackage{fullpage}
\usepackage[tbtags]{amsmath}
\usepackage{amssymb}
\usepackage{mathrsfs}

\newcommand{\nc}[1]{\widehat{#1}}

\begin{document}

%%%%%%%%%%%%%%%%%%%%%%%%%%%%%%%%%%%%%%%%%%%%%%%%%%%%%%%%%%%%%%%%%%%%%%%%%%%%%%%

\rightline{\small{\textit{Phys.~Rev.} D 71 (2005) 045013}}

\bigskip\bigskip
{\huge
\centerline{\textbf{Maps for currents and anomalies}}
\centerline{\textbf{in noncommutative gauge theories}}
}

\bigskip\bigskip
\centerline{\textbf{Rabin Banerjee and Kuldeep Kumar}}

\medskip
\centerline{\textit{S. N. Bose National Centre for Basic Sciences,}}
\centerline{\textit{JD Block, Sector 3, Salt Lake, Kolkata 700098, India}}

\smallskip
\centerline{E-mail: \texttt{rabin@bose.res.in}, \texttt{kuldeep@bose.res.in}}
\date{}

\bigskip
\begin{abstract}
We derive maps relating currents and their divergences in non-Abelian $\mathrm{U}(N)$ noncommutative gauge theory with the corresponding expressions in the ordinary (commutative) description. For the $\mathrm{U}(1)$ theory, in the slowly-varying-field approximation, these maps are also seen to connect the star-gauge-covariant anomaly in the noncommutative theory with the standard Adler--Bell--Jackiw anomaly in the commutative version. For arbitrary fields, derivative corrections to the maps are explicitly computed up to $O(\theta^{2})$.\\[4pt]
\end{abstract}

\bigskip

%%%%%%%%%%%%%%%%%%%%%%%%%%%%%%%%%%%%%%%%%%%%%%%%%%%%%%%%%%%%%%%%%%%%%%%%%%%%%%%

\section{Introduction}

The Seiberg--Witten (SW) map \cite{SW} ensures the stability of classical gauge transformations for theories defined on noncommutative and usual (commutative) spacetime. The field redefinition contained in this map thus provides an alternative method of studying noncommutative gauge theories by recasting these in terms of their commutative equivalents. Maps for the matter sector \cite{BGPSW, Riv, Yan, BY} as well as for currents and energy--momentum tensors \cite{BLY} have also been provided.

An intriguing issue is the validity of such classical maps at the quantum level. Studies in this direction \cite{JSW, KOO, Kam} have principally focussed on extending the purported classical equivalence of Chern--Simons theories (in $2+1$ dimensions) in different descriptions \cite{GS, Ban} to the quantum formulation.

In this paper, we provide an alternative approach to study these quantum aspects by relating the current-divergence anomalies in the noncommutative and commutative pictures through a SW-type map. Taking a cue from an earlier analysis involving one of us \cite{BLY}, we first derive a map connecting the star-gauge-covariant current in the noncommutative gauge theory with the gauge-invariant current in the $\theta$-expanded gauge theory, where $\theta$ is the noncommutativity parameter. From this relation, a mapping between the (star-) covariant divergence of the covariant current and the ordinary divergence of the invariant current in the two descriptions, respectively, is deduced. We find that ordinary current-conservation in the $\theta$-expanded theory implies covariant conservation in the original noncommutative theory, and vice versa. The result is true irrespective of the choice of the current to be vector or axial vector. This is also to be expected on classical considerations.

The issue is quite non-trivial for a quantum treatment due to the occurrence of current-divergence anomalies for axial (chiral) currents. Since the star-gauge-covariant anomaly is known \cite{AS, GM} and the gauge-invariant anomaly in the $\theta$-expanded theory is also known,\footnote{This is in fact identical to the ordinary Adler--Bell--Jackiw (ABJ) anomaly \cite{BMR}.} it is possible to test the map by inserting these expressions. We find that the classical map does not hold in general. However, if we confine to a slowly-varying-field approximation,\footnote{This approximation is also used in Ref.~\cite{SW} to show the equivalence of Dirac--Born--Infeld (DBI) actions in the two descriptions.} then there is a remarkable set of simplifications and the classical map holds. We also give a modified map, that includes the derivative corrections, which is valid for arbitrary field configurations.

The paper is organized as follows. After briefly summarizing the standard SW map in Sec.~\ref{sec:SW}, the map for currents and their divergences is derived in Sec.~\ref{sec:map}. Here the treatment is for the non-Abelian gauge group $\mathrm{U}(N)$. In Sec.~\ref{sec:Abelian}, we discuss the map for anomalous currents and their divergences. The Abelian $\mathrm{U}(1)$ theory is considered and results are given up to $O(\theta^{2})$. As already mentioned, the map for the axial anomalies (in two and four dimensions) holds in the slowly-varying-field limit. A possible scheme is discussed in Sec.~\ref{sec:higher}, whereby further higher-order results are confirmed. Especially, $O(\theta^{3})$ computations are done in some detail. Our concluding remarks are given in Sec.~\ref{sec:conc} where we also briefly discuss the implications of this analysis on the definition of effective actions.

%%%%%%%%%%%%%%%%%%%%%%%%%%%%%%%%%%%%%%%%%%%%%%%%%%%%%%%%%%%%%%%%%%%%%%%%%%%%%%%

\section{\label{sec:SW}A brief review of the Seiberg--Witten map}

Let us begin by briefly reviewing the salient features of the SW map. The ordinary Yang--Mills action is given by
\begin{equation}\label{S-YM}
S_{\mathrm{YM}} = -\frac{1}{4}\int\!\mathrm{d}^{4}x\,\mathrm{Tr}\left(F_{\mu\nu}F^{\mu\nu}\right),
\end{equation}
where the non-Abelian field strength is defined as
\begin{equation}\label{F.d}
F_{\mu\nu}=\partial_{\mu}A_{\nu}-\partial_{\nu}A_{\mu}-\mathrm{i}[A_{\mu},A_{\nu}]
\end{equation}
in terms of the Hermitian U($N$) gauge fields $A_{\mu}(x)$. The noncommutativity of spacetime is characterized by the algebra
\begin{equation}\label{theta.d}
\left[x^{\alpha},x^{\beta}\right]_{\star} \equiv x^{\alpha}\star x^{\beta}-x^{\beta}\star x^{\alpha} = \mathrm{i}\theta^{\alpha\beta},
\end{equation}
where the noncommutativity parameter $\theta^{\alpha\beta}$ is real and antisymmetric. The star product of two fields $A(x)$ and $B(x)$ is defined as
\begin{equation}\label{star.d}
(A\star B)(x)=\left.\exp\left(\frac{\mathrm{i}}{2}\theta^{\alpha\beta}\partial_{\alpha}\partial'_{\beta}\right)A(x)B(x')\right|_{x'=x},
\end{equation}
where $\partial'_{\beta}\equiv \frac{\partial}{\partial x'^{\beta}}$. In noncommutative spacetime, the usual multiplication of functions is replaced by the star product. The Yang--Mills theory is generalized to
\begin{equation}\label{S-hat-YM}
\nc{S}_{\mathrm{YM}} = -\frac{1}{4}\int\!\mathrm{d}^{4}x\,\mathrm{Tr}\left(\nc{F}_{\mu\nu}\star\nc{F}^{\mu\nu}\right)
\end{equation}
with the noncommutative field strength
\begin{equation}\label{F-hat.d}
\nc{F}_{\mu\nu}=\partial_{\mu}\nc{A}_{\nu}-\partial_{\nu}\nc{A}_{\mu}-\mathrm{i}\left[\nc{A}_{\mu},\nc{A}_{\nu}\right]_{\star}.
\end{equation}
This theory reduces to the conventional U($N$) Yang--Mills theory for $\theta \rightarrow 0$.

To first order in $\theta$, it is possible to relate the variables in the noncommutative spacetime with those in the usual one by the classical maps \cite{SW}
\begin{gather}
\label{A-hat.m}
\nc{A}_{\mu} = A_{\mu}-\frac{1}{4}\theta^{\alpha\beta}\{A_{\alpha},\partial_{\beta}A_{\mu}+F_{\beta\mu}\}+O(\theta^{2}),\\
\label{F-hat.m}
\nc{F}_{\mu\nu} = F_{\mu\nu}+\frac{1}{4}\theta^{\alpha\beta}\left(2\{F_{\mu\alpha},F_{\nu\beta}\}-\{A_{\alpha},\mathrm{D}_{\beta}F_{\mu\nu}+\partial_{\beta}F_{\mu\nu}\}\right)+O(\theta^{2}),
\end{gather}
where $\{\cdots,\cdots\}$ appearing on the right-hand sides stands for the anticommutator and $\mathrm{D}_{\beta}$ denotes the covariant derivative defined as $\mathrm{D}_{\beta}\lambda=\partial_{\beta}\lambda+\mathrm{i}[\lambda,A_{\beta}]$. A further map among gauge parameters,
\begin{equation}\label{lmda-hat.m}
\nc{\lambda}=\lambda+\frac{1}{4}\theta^{\alpha\beta}\left\{\partial_{\alpha}\lambda,A_{\beta}\right\}+O(\theta^{2}),
\end{equation}
ensures the stability of gauge transformations
\begin{gather}
\label{A.gt}
\nc{\delta}_{\nc{\lambda}}\nc{A}_{\mu} = \partial_{\mu}\nc{\lambda}+\mathrm{i}\left[\nc{\lambda},\nc{A}_{\mu}\right]_{\star} \equiv \nc{\mathrm{D}}_{\mu}\star\nc{\lambda},\\
\label{A-hat.gt}
\delta_{\lambda}A_{\mu} = \partial_{\mu}\lambda+\mathrm{i}\left[\lambda,A_{\mu}\right] \equiv \mathrm{D}_{\mu}\lambda.
\end{gather}
That is, if two ordinary gauge fields $A_{\mu}$ and $A'_{\mu}$ are equivalent by an ordinary gauge transformation, then the corresponding noncommutative gauge fields, $\nc{A}_{\mu}$ and $\nc{A'}_{\mu}$, will also be gauge-equivalent by a noncommutative gauge transformation. It may be noted that the map (\ref{F-hat.m}) is a consequence of the map (\ref{A-hat.m}) following from the definition (\ref{F-hat.d}) of the noncommutative field strength. The field strengths $F_{\mu\nu}$ and $\nc{F}_{\mu\nu}$ transform covariantly under the usual and the star-gauge transformations, respectively:
\begin{equation}\label{F&F-hat.gt}
\delta_{\lambda}F_{\mu\nu} = \mathrm{i}\left[\lambda,F_{\mu\nu}\right],\quad
\nc{\delta}_{\nc{\lambda}}\nc{F}_{\mu\nu} = \mathrm{i}\left[\nc{\lambda},\nc{F}_{\mu\nu}\right]_{\star}.
\end{equation}

The gauge fields $A_{\mu}(x)$ may be expanded in terms of the Lie-algebra generators $T^{a}$ of U($N$) as $A_{\mu}(x)=A_{\mu}^{a}(x)T^{a}$. These generators satisfy
\begin{equation}\label{T.p}
\left[T^{a},T^{b}\right] = \mathrm{i}f^{abc}T^{c}, \quad \left\{T^{a},T^{b}\right\}=d^{abc}T^{c}, \quad \mathrm{Tr}\left(T^{a}T^{b}\right)=\delta^{ab}.
\end{equation}
We shall take the structure functions $f^{abc}$ and $d^{abc}$ to be, respectively, totally antisymmetric and totally symmetric. The Yang--Mills action (\ref{S-YM}) can now be rewritten as\footnote{A lower gauge index is equivalent to a raised one---whether a gauge index appears as a superscript or as a subscript is a matter of notational convenience.}
\begin{equation}\label{S-YM2}
S_{\mathrm{YM}} = -\frac{1}{4}\int\!\mathrm{d}^{4}x\,F_{\mu\nu}^{a}F^{\mu\nu}_{a},
\end{equation}
where
\begin{equation}\label{F.d2}
F_{\mu\nu}^{a}=\partial_{\mu}A_{\nu}^{a}-\partial_{\nu}A_{\mu}^{a}+f^{abc}A_{\mu}^{b}A_{\nu}^{c}.
\end{equation}
In view of relations (\ref{T.p}), the maps (\ref{A-hat.m})--(\ref{lmda-hat.m}) can also be written as
\begin{gather}
\label{A-hat.m2}
\nc{A}_{\mu}^{c} = A_{\mu}^{c}-\frac{1}{4}\theta^{\alpha\beta}d^{abc}A_{\alpha}^{a}\left(\partial_{\beta}A_{\mu}^{b}+F_{\beta\mu}^{b}\right)+O(\theta^{2}),\\
\label{F-hat.m2}
\nc{F}_{\mu\nu}^{c} = F_{\mu\nu}^{c}+\frac{1}{2}\theta^{\alpha\beta}d^{abc}\left(F_{\mu\alpha}^{a}F_{\nu\beta}^{b}-A_{\alpha}^{a}\partial_{\beta}F_{\mu\nu}^{b}+\frac{1}{2}f^{bde}A_{\alpha}^{a}A_{\beta}^{e}F_{\mu\nu}^{d}\right)+O(\theta^{2}),\\
\label{lmda-hat.m2}
\nc{\lambda}^{c} = \lambda^{c}+\frac{1}{4}\theta^{\alpha\beta}d^{abc}\partial_{\alpha}\lambda^{a}A_{\beta}^{b}+O(\theta^{2}),
\end{gather}
and the gauge transformations (\ref{A.gt})--(\ref{F&F-hat.gt}) as
\begin{gather}
\label{A.gt2}
\delta_{\lambda}A_{\mu}^{a} = \partial_{\mu}\lambda^{a}+f^{abc}A_{\mu}^{b}\lambda^{c},\\
\label{F.gt2}
\delta_{\lambda}F_{\mu\nu}^{a} = f^{abc}F_{\mu\nu}^{b}\lambda^{c},\\
\label{Ahat.gt2}\begin{split}
\nc{\delta}_{\nc{\lambda}}\nc{A}_{\mu}^{a} &= \partial_{\mu}\nc{\lambda}^{a}+\frac{\mathrm{i}}{2}d^{abc}\left[\nc{\lambda}^{b},\nc{A}_{\mu}^{c}\right]_{\star}-\frac{1}{2}f^{abc}\left\{\nc{\lambda}^{b},\nc{A}_{\mu}^{c}\right\}_{\star}\\
&= \partial_{\mu}\nc{\lambda}^{a}+f^{abc}\nc{A}_{\mu}^{b}\nc{\lambda}^{c}+\frac{1}{2}\theta^{\alpha\beta}d^{abc}\partial_{\alpha}\nc{A}_{\mu}^{b}\partial_{\beta}\nc{\lambda}^{c}+O(\theta^{2}),\end{split}\\
\label{Fhat.gt2}\begin{split}
\nc{\delta}_{\nc{\lambda}}\nc{F}_{\mu\nu}^{a} &= \frac{\mathrm{i}}{2}d^{abc}\left[\nc{\lambda}^{b},\nc{F}_{\mu\nu}^{c}\right]_{\star}-\frac{1}{2}f^{abc}\left\{\nc{\lambda}^{b},\nc{F}_{\mu\nu}^{c}\right\}_{\star}\\
&= f^{abc}\nc{F}_{\mu\nu}^{b}\nc{\lambda}^{c}+\frac{1}{2}\theta^{\alpha\beta}d^{abc}\partial_{\alpha}\nc{F}_{\mu\nu}^{b}\partial_{\beta}\nc{\lambda}^{c}+O(\theta^{2}).
\end{split}
\end{gather}

%%%%%%%%%%%%%%%%%%%%%%%%%%%%%%%%%%%%%%%%%%%%%%%%%%%%%%%%%%%%%%%%%%%%%%%%%%%%%%%

\section{\label{sec:map}The map for non-Abelian currents: classical aspects}

In order to discuss noncommutative gauge theories with sources, it is essential to have a map for the sources also, so that a complete transition between noncommutative gauge theories and the usual ones is possible. Such a map was first briefly discussed in Ref.~\cite{BLY} for the Abelian case. We consider the non-Abelian case in this section.

Let the noncommutative action be defined as
\begin{equation}\label{S-hat}
\nc{S}(\nc{A},\nc{\psi})=\nc{S}_{\mathrm{YM}}(\nc{A})+\nc{S}_{\mathrm{M}}(\nc{\psi},\nc{A}),
\end{equation}
where $\nc{\psi}_{\alpha}$ are the charged matter fields. The equation of motion for $\nc{A}_{\mu}^{a}$ is\footnote{We mention that the noncommutative gauge field $\nc{A}_{\mu}$ is in general an element of the enveloping algebra of the gauge group. Only for specific cases, as for instance the considered case of $\mathrm{U}(N)$ gauge symmetry, it is Lie-algebra valued.}
\begin{equation}\label{eom-nc}
\frac{\delta\nc{S}_{\mathrm{YM}}}{\delta\nc{A}_{\mu}^{a}}=\nc{\mathrm{D}}_{\nu}\star\nc{F}^{\nu\mu}_{a}=-\nc{J}^{\mu}_{a},
\end{equation}
where
\begin{equation}\label{J-hat.d}
\nc{J}^{\mu}_{a}=\left.\frac{\delta\nc{S}_{\mathrm{M}}}{\delta\nc{A}_{\mu}^{a}}\right|_{\nc{\psi}}.
\end{equation}
Equation (\ref{eom-nc}) shows that $\nc{J}^{\mu}_{a}$ transforms covariantly under the star-gauge transformation:
\begin{equation}\label{sgt2}
\nc{\delta}_{\nc{\lambda}}\nc{J}^{\mu} = -\mathrm{i}\left[\nc{J}^{\mu},\nc{\lambda}\right]_{\star}, \quad
\nc{\delta}_{\nc{\lambda}}\nc{J}^{\mu}_{a} = f^{abc}\nc{J}^{\mu}_{b}\nc{\lambda}^{c}+\frac{1}{2}\theta^{\alpha\beta}d^{abc}\partial_{\alpha}\nc{J}^{\mu}_{b}\partial_{\beta}\nc{\lambda}^{c}+O(\theta^{2}).
\end{equation}
Also, it satisfies the noncommutative covariant conservation law
\begin{equation}\label{DhJh3}
\nc{\mathrm{D}}_{\mu}\star\nc{J}^{\mu}_{a}=0,
\end{equation}
which may be seen from Eq.~(\ref{eom-nc}) by taking the noncommutative covariant divergence.

The use of SW map in the action (\ref{S-hat}) gives its $\theta$-expanded version in commutative space:
\begin{equation}\label{S-theta}
\nc{S}(\nc{A},\nc{\psi})\rightarrow S^{\theta}(A,\psi)=S^{\theta}_{\mathrm{YM}}(A)+S^{\theta}_{\mathrm{M}}(\psi,A),
\end{equation}
where $S^{\theta}_{\mathrm{YM}}(A)$ contains all terms involving $A_{\mu}^{a}$ only, and is given by
\begin{equation}\label{S-th-YM}
S^{\theta}_{\mathrm{YM}} = -\frac{1}{4}\int\!\mathrm{d}^{4}x\left[F_{\mu\nu}^{a}F^{\mu\nu}_{a}+\theta^{\alpha\beta}d^{abc}F^{\mu\nu}_{a}\left(F_{\mu\alpha}^{b}F_{\nu\beta}^{c}+\frac{1}{4}F_{\beta\alpha}^{b}F_{\mu\nu}^{c}\right)+O(\theta^{2})\right],
\end{equation}
and we have dropped a boundary term in order to express it solely in terms of the field strength. The equation of motion following from the action (\ref{S-theta}) is
\begin{equation}\label{eom-th}
\frac{\delta S^{\theta}_{\mathrm{YM}}}{\delta A_{\mu}^{a}}=-J^{\mu}_{a},
\end{equation}
where
\begin{equation}\label{J.d}
J^{\mu}_{a}=\left.\frac{\delta S^{\theta}_{\mathrm{M}}}{\delta A_{\mu}^{a}}\right|_{\psi}.
\end{equation}
Expectedly, from these relations, it follows that $J^{\mu}_{a}$ transforms covariantly,
\begin{equation}\label{gt6}
\delta_{\lambda}J^{\mu} = -\mathrm{i}\left[J^{\mu}, \lambda\right], \quad \delta_{\lambda}J^{\mu}_{a} = f^{abc}J^{\mu}_{b}\lambda^{c},
\end{equation}
and satisfies the covariant conservation law
\begin{equation}\label{DJ3}
\mathrm{D}_{\mu}J^{\mu}_{a} = 0.
\end{equation}

Now the application of SW map on the right-hand side of Eq.~(\ref{J-hat.d}) yields the relation between $\nc{J}^{\mu}_{a}$ and $J^{\mu}_{a}$:
\begin{equation}\label{J-hat.m}
\nc{J}^{\mu}_{a}(x) = \int\!\mathrm{d}^{4}y\left[\left.\frac{\delta S^{\theta}_{\mathrm{M}}}{\delta A_{\nu}^{c}(y)}\right|_{\psi}\frac{\delta A_{\nu}^{c}(y)}{\delta \nc{A}_{\mu}^{a}(x)}+\left.\frac{\delta S^{\theta}_{\mathrm{M}}}{\delta \psi_{\alpha}^{c}(y)}\right|_{A}\frac{\delta \psi_{\alpha}^{c}(y)}{\delta \nc{A}_{\mu}^{a}(x)}\right] = \int\!\mathrm{d}^{4}y\,J^{\nu}_{c}(y)\frac{\delta A_{\nu}^{c}(y)}{\delta \nc{A}_{\mu}^{a}(x)},
\end{equation}
where the second term obtained in the first step has been dropped on using the equation of motion for $\psi_{\alpha}^{a}$.

We consider Eq.~(\ref{J-hat.m}) as a closed form for the map among the sources. To get its explicit structure, the map (\ref{A-hat.m2}) among the gauge potentials is necessary. Since the map (\ref{A-hat.m2}) is a classical result, the map for the sources obtained in this way is also classical.

Let us next obtain the explicit form of this map up to first order in $\theta$. Using the map (\ref{A-hat.m2}) and its inverse,
\begin{equation}\label{A-in.m2}
A_{\mu}^{c}=\nc{A}_{\mu}^{c}+\frac{1}{4}\theta^{\alpha\beta}d^{abc}\nc{A}_{\alpha}^{a}\left(\partial_{\beta}\nc{A}_{\mu}^{b}+\nc{F}_{\beta\mu}^{b}\right)+O(\theta^{2}),
\end{equation}
we can compute the functional derivative
\begin{equation}\label{fund}\begin{split}
\frac{\delta A_{\nu}^{c}(y)}{\delta \nc{A}_{\mu}^{a}(x)} &= \delta^{\mu}_{\nu}\delta^{ac}\delta(x-y)\\
&\quad {}+\frac{1}{4}\theta^{\alpha\beta}\delta^{\mu}_{\nu}\left[2d^{abc}A_{\alpha}^{b}(y)\partial_{\beta}^{y}\delta(x-y)+d^{edc}f^{bad}A_{\alpha}^{e}(y)A_{\beta}^{b}(y)\delta(x-y)\right]\\
&\quad {}-\frac{1}{4}\theta^{\alpha\mu}\left[d^{abc}A_{\alpha}^{b}(y)\partial_{\nu}^{y}\delta(x-y)\right.\\
&\qquad\qquad\quad \left.{}+\left(d^{abc}\partial_{\alpha}^{y}A_{\nu}^{b}(y)+d^{abc}F_{\alpha\nu}^{b}(y)-d^{edc}f^{dab}A_{\alpha}^{e}(y)A_{\nu}^{b}(y)\right)\delta(x-y)\right]\\
&\quad {}+O(\theta^{2}),\end{split}
\end{equation}
where $\partial_{\beta}^{y}$ stands for $\frac{\partial}{\partial y^{\beta}}$. Putting this in Eq.~(\ref{J-hat.m}), we get
\begin{equation}\label{J-hat.m2'}\begin{split}
\nc{J}^{\mu}_{a} &= J^{\mu}_{a}-\frac{1}{2}\theta^{\alpha\beta}\left[d^{abc}\partial_{\beta}\left(A_{\alpha}^{b}J^{\mu}_{c}\right)-\frac{1}{2}d^{edc}f^{bad}A_{\alpha}^{e}A_{\beta}^{b}J^{\mu}_{c}\right]\\
&\quad {}-\frac{1}{2}\theta^{\alpha\mu}\left[d^{abc}F_{\alpha\nu}^{b}J^{\nu}_{c}-\frac{1}{2}\left(d^{cad}f^{dbe}+d^{bcd}f^{dae}\right)A_{\alpha}^{b}A_{\nu}^{e}J^{\nu}_{c}-\frac{1}{2}d^{abd}A_{\alpha}^{b}\partial_{\nu}J^{\nu}_{d}\right]+O(\theta^{2}).\end{split}
\end{equation}
Since $\mathrm{D}_{\nu}J^{\nu}_{a} \equiv \partial_{\nu}J^{\nu}_{a}-f^{abc}J^{\nu}_{b}A_{\nu}^{c}$, we can use Eq.~(\ref{DJ3}) to substitute
\begin{equation}\label{xxx}
\partial_{\nu}J^{\nu}_{d}=f^{dce}J^{\nu}_{c}A_{\nu}^{e}
\end{equation}
in the last term on the right-hand side of Eq.~(\ref{J-hat.m2'}) to obtain
\begin{equation}\label{J-hat.m2}
\nc{J}^{\mu}_{a}=J^{\mu}_{a}-\frac{1}{2}\theta^{\alpha\beta}\left[d^{abc}\partial_{\beta}\left(A_{\alpha}^{b}J^{\mu}_{c}\right)-\frac{1}{2}d^{edc}f^{bad}A_{\alpha}^{e}A_{\beta}^{b}J^{\mu}_{c}\right]
-\frac{1}{2}\theta^{\alpha\mu}d^{abc}F_{\alpha\nu}^{b}J^{\nu}_{c}+O(\theta^{2}),
\end{equation}
where we have used the identity
\begin{equation}\label{df.i}
d^{abd}f^{dce}+d^{bcd}f^{dae}+d^{cad}f^{dbe}=0.
\end{equation}
As a simple yet non-trivial consistency check, we show the stability of the map under gauge transformations. Under the ordinary gauge transformations given by Eqs.~(\ref{A.gt2}) and (\ref{F.gt2}), and using the covariant transformation law (\ref{gt6}) for $J^{\mu}_{a}$, the right-hand side of Eq.~(\ref{J-hat.m2}) transforms as
\begin{equation}\label{J-hat.gt1}\begin{split}
\delta_{\lambda}\nc{J}^{\mu}_{a} &= f^{abc}J^{\mu}_{b}\lambda^{c}
-\frac{1}{2}\theta^{\alpha\beta}\bigg[d^{abc}\partial_{\beta}J^{\mu}_{c}\partial_{\alpha}\lambda^{b}+d^{cdb}f^{bea}\partial_{\beta}\left(A_{\alpha}^{d}J^{\mu}_{c}\lambda^{e}\right)\\
&\qquad\qquad\qquad\qquad\quad {}+\frac{1}{2}\left(d^{ecb}f^{bda}-d^{cdb}f^{bea}\right)A_{\alpha}^{d}J^{\mu}_{c}\partial_{\beta}\lambda^{e}\\
&\qquad\qquad\qquad\qquad\quad {}+\frac{1}{2}d^{gcd}\left(f^{abe}f^{edh}+f^{dae}f^{ebh}\right)A_{\alpha}^{g}A_{\beta}^{b}J^{\mu}_{c}\lambda^{h}\bigg]\\
&\quad {}+\frac{1}{2}\theta^{\alpha\mu}d^{cdb}f^{bae}F_{\alpha\nu}^{d}J^{\nu}_{c}\lambda^{e}+O(\theta^{2}),\end{split}
\end{equation}
where we have used the relation (\ref{df.i}). On the other hand, using the maps (\ref{lmda-hat.m2}) and (\ref{J-hat.m2}), and the identity
\begin{equation}\label{ff.i}
f^{abe}f^{edh}+f^{bde}f^{eah}+f^{dae}f^{ebh}=0,
\end{equation}
the right-hand side of the second relation in Eq.~(\ref{sgt2}) reproduces the right-hand side of Eq.~(\ref{J-hat.gt1}). Hence,
\begin{equation}\label{J-hat.gt2}
\nc{\delta}_{\nc{\lambda}}\nc{J}^{\mu}_{a} = \delta_{\lambda}\nc{J}^{\mu}_{a},
\end{equation}
thereby proving the stability of the map (\ref{J-hat.m2}) under the gauge transformations. This statement is equivalent to the usual notion of stability which ensures that the star-gauge-transformed noncommutative current is mapped to the usual-gauge-transformed ordinary current, as may be verified by performing a Taylor expansion of the right-hand side of $\nc{J}^{\mu}_{a}(J, A)+\nc{\delta}_{\nc{\lambda}}\nc{J}^{\mu}_{a}(J, A) = \nc{J}^{\mu}_{a}(J+\delta_{\lambda}J, A+\delta_{\lambda}A)$ and comparing both sides.\footnote{Exactly the same thing happens when discussing the stability of the map (\ref{A-hat.m}) for the potentials.} 

It is worthwhile to mention that the use  of Eq.~(\ref{xxx}) in obtaining the map (\ref{J-hat.m2}) is crucial to get the correct transformation property of $\nc{J}^{\mu}_{a}$. This is because issues of gauge covariance and covariant conservation are not independent. In an ordinary Abelian gauge theory, for example, current conservation and gauge invariance are related. Likewise, in the non-Abelian case, covariant conservation and gauge covariance are related. This intertwining property is a peculiarity of the mapping among the sources and is not to be found in the mapping among the potentials or the field strengths. 

From these results, it is possible to give a map for the covariant derivatives of the currents. Recall that
\begin{equation}\label{D-hJ-h}\begin{split}
\nc{\mathrm{D}}_{\mu}\star\nc{J}^{\mu}_{a} &= \partial_{\mu}\nc{J}^{\mu}_{a}+\frac{\mathrm{i}}{2}d^{abc}\left[\nc{J}^{\mu}_{b},\nc{A}_{\mu}^{c}\right]_{\star}-\frac{1}{2}f^{abc}\left\{\nc{J}^{\mu}_{b},\nc{A}_{\mu}^{c}\right\}_{\star}\\
&= \partial_{\mu}\nc{J}^{\mu}_{a}+f^{abc}\nc{A}_{\mu}^{b}\nc{J}^{\mu}_{c}+\frac{1}{2}\theta^{\alpha\beta}d^{abc}\partial_{\alpha}\nc{A}_{\mu}^{b}\partial_{\beta}\nc{J}^{\mu}_{c}+O(\theta^{2}),\end{split}
\end{equation}
which, using the maps (\ref{A-hat.m2}) and (\ref{J-hat.m2}), gives
\begin{equation}\label{D-hJ-h2}
\nc{\mathrm{D}}_{\mu}\star\nc{J}^{\mu}_{a} = \mathrm{D}_{\mu}J^{\mu}_{a}-\frac{1}{2}\theta^{\alpha\beta}\left[d^{abc}\partial_{\beta}\left(A_{\alpha}^{b}\mathrm{D}_{\mu}J^{\mu}_{c}\right)-\frac{1}{2}d^{edc}f^{bad}A_{\alpha}^{e}A_{\beta}^{b}\mathrm{D}_{\mu}J^{\mu}_{c}\right]+O(\theta^{2}),
\end{equation}
where we have used the Jacobi identities (\ref{df.i}) and (\ref{ff.i}), and the relation (\ref{xxx}). Thus we see that covariant conservation of the ordinary current, $\mathrm{D}_{\mu}J^{\mu}_{a}=0$, implies that $\nc{J}^{\mu}_{a}$ given by Eq.~(\ref{J-hat.m2}) indeed satisfies the noncommutative covariant conservation law, $\nc{\mathrm{D}}_{\mu}\star\nc{J}^{\mu}_{a}=0$. This is also to be expected from classical notions.

At this point, an intriguing issue arises. Is it possible to use Eq.~(\ref{D-hJ-h2}) to relate the anomalies in the different descriptions? Indeed the analysis presented for the vector current can be readily taken over for the chiral current. Classically everything would be fine since the relevant currents are both conserved. At the quantum level, however, the chiral currents are not conserved. We would like to ascertain whether the relation (\ref{D-hJ-h2}) is still valid by substituting the relevant chiral anomalies in place of $\nc{\mathrm{D}}_{\mu}\star\nc{J}^{\mu}_{a}$ and $\mathrm{D}_{\mu}J^{\mu}_{a}$. Since the main aspects get highlighted for the Abelian theory itself, we confine to this case, and present a detailed analysis in the remainder of this paper.

%%%%%%%%%%%%%%%%%%%%%%%%%%%%%%%%%%%%%%%%%%%%%%%%%%%%%%%%%%%%%%%%%%%%%%%%%%%%%%%

\section{The Abelian case: classical and quantum aspects\label{sec:Abelian}}

Some discussion on the use of the map (\ref{D-hJ-h2}) (in the Abelian case) for relating anomalies (up to $O(\theta)$) was earlier given in Ref.~\cite{BLY}. In order to gain a deeper understanding, it is essential to consider higher orders in $\theta$. Keeping this in mind, we present a calculation up to $O(\theta^{2})$ for two- and four-dimensional theories.

The maps to the second order in $\theta$ in the Abelian case are given by \cite{Fid}
\begin{gather}
\label{101}\begin{split}
\nc{A}_{\mu} &= A_{\mu}-\frac{1}{2}\theta^{\alpha\beta}A_{\alpha}\left(\partial_{\beta}A_{\mu}+F_{\beta\mu}\right)\\
&\quad {}+\frac{1}{6}\theta^{\alpha\beta}\theta^{\kappa\sigma}A_{\alpha}\left[\partial_{\beta}\left(A_{\kappa}\partial_{\sigma}A_{\mu}+2A_{\kappa}F_{\sigma\mu}\right)+F_{\beta\kappa}\left(\partial_{\sigma}A_{\mu}+2F_{\sigma\mu}\right)\right]+O(\theta^{3}),\end{split}\\
\label{107}\begin{split}
\nc{F}_{\mu\nu} &= F_{\mu\nu}-\theta^{\alpha\beta}\left(A_{\alpha}\partial_{\beta}F_{\mu\nu}+F_{\mu\alpha}F_{\beta\nu}\right)\\
&\quad {}+\frac{1}{2}\theta^{\alpha\beta}\theta^{\kappa\sigma}\left[A_{\alpha}\partial_{\beta}\left(A_{\kappa}\partial_{\sigma}F_{\mu\nu}+2F_{\mu\kappa}F_{\sigma\nu}\right)+F_{\beta\kappa}\left(A_{\alpha}\partial_{\sigma}F_{\mu\nu}+2F_{\mu\alpha}F_{\sigma\nu}\right)\right]\\
&\quad {}+O(\theta^{3}),\end{split}\\
\label{102}
\nc{\lambda} = \lambda-\frac{1}{2}\theta^{\alpha\beta}A_{\alpha}\partial_{\beta}\lambda+\frac{1}{6}\theta^{\alpha\beta}\theta^{\kappa\sigma}A_{\alpha}\left[\partial_{\beta}\left(A_{\kappa}\partial_{\sigma}\lambda\right)+F_{\beta\kappa}\partial_{\sigma}\lambda\right]+O(\theta^{3}),
\end{gather}
which ensure the stability of gauge transformations
\begin{gather}
\label{1040}
\nc{\delta}_{\nc{\lambda}}\nc{A}_{\mu} = \nc{\mathrm{D}}_{\mu}\star\nc{\lambda}
\equiv\partial_{\mu}\nc{\lambda}+\mathrm{i}\left[\nc{\lambda},\nc{A}_{\mu}\right]_{\star}
= \partial_{\mu}\nc{\lambda}+\theta^{\alpha\beta}\partial_{\alpha}\nc{A}_{\mu}\partial_{\beta}\nc{\lambda}+O(\theta^{3}),\\
\label{104}
\delta_{\lambda}A_{\mu} = \partial_{\mu}\lambda.
\end{gather}
Analogous to the non-Abelian theory, the map for currents is consistent with the requirements that while the current $J^{\mu}$ is gauge-invariant and satisfies the ordinary conservation law, $\partial_{\mu}J^{\mu}=0$, the current $\nc{J}^{\mu}$ is star-gauge-covariant and satisfies the noncommutative covariant conservation law, $\nc{\mathrm{D}}_{\mu}\star\nc{J}^{\mu}=0$. Now the currents $J^{\mu}$ and $\nc{J}^{\mu}$ are related by the Abelian version of Eq.~(\ref{J-hat.m}) \cite{BLY},
\begin{equation}\label{109}
\nc{J}^{\mu}(x) = \int\!\mathrm{d}^{4}yJ^{\nu}(y)\frac{\delta A_{\nu}(y)}{\delta\nc{A}_{\mu}(x)},
\end{equation}
which, using the map (\ref{101}) and its inverse,
\begin{equation}\label{110}\begin{split}
A_{\mu} &= \nc{A}_{\mu}+\frac{1}{2}\theta^{\alpha\beta}\nc{A}_{\alpha}\left(\partial_{\beta}\nc{A}_{\mu}+\nc{F}_{\beta\mu}\right)\\
&\quad {}+\frac{1}{6}\theta^{\alpha\beta}\theta^{\kappa\sigma}\nc{A}_{\alpha}\left[\frac{1}{2}\partial_{\beta}\left(\nc{A}_{\kappa}\partial_{\sigma}\nc{A}_{\mu}-\nc{A}_{\kappa}\nc{F}_{\sigma\mu}\right)+\frac{1}{2}\nc{F}_{\beta\kappa}\left(\partial_{\sigma}\nc{A}_{\mu}+5\nc{F}_{\sigma\mu}\right)\right.\\
&\qquad\qquad\qquad\qquad {}+\left.\frac{3}{2}\left(2\nc{A}_{\kappa}\partial_{\beta}\nc{F}_{\sigma\mu}+\partial_{\beta}\nc{A}_{\kappa}\partial_{\sigma}\nc{A}_{\mu}+\partial_{\beta}\nc{A}_{\kappa}\nc{F}_{\sigma\mu}\right)\right]+O(\theta^{3}),\end{split}
\end{equation}
yields the explicit $O(\theta^{2})$ form of the source map:
\begin{equation}\label{111}\begin{split}
\nc{J}^{\mu} &= J^{\mu}-\theta^{\alpha\beta}\left(A_{\alpha}\partial_{\beta}J^{\mu}-\frac{1}{2}F_{\alpha\beta}J^{\mu}\right)+\theta^{\mu\alpha}F_{\alpha\beta}J^{\beta}\\
&\quad {}+\frac{1}{2}\theta^{\alpha\beta}\theta^{\kappa\sigma}\partial_{\alpha}\left(A_{\kappa}F_{\beta\sigma}J^{\mu}-A_{\beta}A_{\kappa}\partial_{\sigma}J^{\mu}+\frac{1}{2}A_{\beta}F_{\kappa\sigma}J^{\mu}\right)-\theta^{\alpha\beta}\theta^{\kappa\mu}\partial_{\alpha}\left(A_{\beta}F_{\kappa\nu}J^{\nu}\right)\\
&\quad {}+O(\theta^{3}),\end{split}
\end{equation}
where we have used $\partial_{\mu}J^{\mu}=0$ to simplify the integrand.\footnote{This is essential to ensure the stability of map (\ref{111}) under appropriate gauge transformations. A similar manipulation was needed for getting the non-Abelian expression (\ref{J-hat.m2}).} The above map, up to $O(\theta)$, was earlier given in Ref.~\cite{BLY}. Now let us check explicitly the stability under the gauge transformations. Under the ordinary gauge transformation, $\delta_{\lambda}A_{\mu}=\partial_{\mu}\lambda$, $\delta_{\lambda}F_{\mu\nu}=0$, and $\delta_{\lambda}J^{\mu}=0$. Hence the right-hand side of Eq.~(\ref{111}) transforms as
\begin{equation}\label{113}\begin{split}
\delta_{\lambda}\nc{J}^{\mu} &= \theta^{\alpha\beta}\partial_{\alpha}J^{\mu}\partial_{\beta}\lambda+\theta^{\alpha\beta}\theta^{\mu\kappa}\partial_{\alpha}\left(F_{\kappa\nu}J^{\nu}\right)\partial_{\beta}\lambda\\
&\quad {}+\frac{1}{2}\theta^{\alpha\beta}\theta^{\kappa\sigma}\left[2\partial_{\beta}\partial_{\sigma}\left(A_{\kappa}J^{\mu}\right)\partial_{\alpha}\lambda-\partial_{\beta}\left(A_{\kappa}\partial_{\sigma}\lambda\right)\partial_{\alpha}J^{\mu}\right]+O(\theta^{3}).\end{split}
\end{equation}
On the other hand,
\begin{equation}\label{1131}
\nc{\delta}_{\nc{\lambda}}\nc{J}^{\mu} = \mathrm{i}\left[\nc{\lambda},\nc{J}^{\mu}\right]_{\star} = \theta^{\alpha\beta}\partial_{\alpha}\nc{J}^{\mu}\partial_{\beta}\nc{\lambda}+O(\theta^{3}).
\end{equation}
Next, using the maps (\ref{102}) and (\ref{111}) in the above equation, one finds that the right-hand side of Eq.~(\ref{113}) is reproduced. Hence,
\begin{equation}\label{11319}
\nc{\delta}_{\nc{\lambda}}\nc{J}^{\mu} = \delta_{\lambda}\nc{J}^{\mu},
\end{equation}
thereby proving the gauge-equivalence, as observed earlier. Furthermore, using the maps (\ref{101}) and (\ref{111}), the covariant divergence of $\nc{J}^{\mu}$,
\begin{equation}\label{114}
\nc{\mathrm{D}}_{\mu}\star\nc{J}^{\mu} = \partial_{\mu}\nc{J}^{\mu}+\mathrm{i}\left[\nc{J}^{\mu},\nc{A}_{\mu}\right]_{\star} = \partial_{\mu}\nc{J}^{\mu}-\theta^{\alpha\beta}\partial_{\alpha}\nc{J}^{\mu}\partial_{\beta}\nc{A}_{\mu}+O(\theta^{3}),
\end{equation}
can be expressed as
\begin{equation}\label{115}\begin{split}
\nc{\mathrm{D}}_{\mu}\star\nc{J}^{\mu} &= \partial_{\mu}J^{\mu}+\theta^{\alpha\beta}\partial_{\alpha}\left(A_{\beta}\partial_{\mu}J^{\mu}\right)
+\frac{1}{2}\theta^{\alpha\beta}\theta^{\kappa\sigma}\partial_{\alpha}\left[A_{\kappa}F_{\beta\sigma}\partial_{\mu}J^{\mu}-A_{\beta}\partial_{\sigma}\left(A_{\kappa}\partial_{\mu}J^{\mu}\right)\right]\\
&\quad {}+O(\theta^{3}),\end{split}
\end{equation}
where each term on the right-hand side involves $\partial_{\mu}J^{\mu}$, so that the covariant conservation of $\nc{J}^{\mu}$ follows from the ordinary conservation of $J^{\mu}$. This is the Abelian analogue of Eq.~(\ref{D-hJ-h2}), but valid up to $O(\theta^{2})$.

We are now in a position to discuss the mapping of anomalies. Since the maps have been obtained for the gauge currents, the anomalies refer to chiral anomalies found in chiral gauge theories. Moreover, we implicitly assume a regularization which preserves vector-current conservation so that the chiral anomaly $\partial_\mu[\bar\psi\gamma^\mu(\frac{1+\gamma_5}{2})\psi]$ is proportional to the usual ABJ anomaly $\partial_{\mu} J^{\mu}_{5}$ \cite{BB}. The first step is to realize that the standard ABJ anomaly \cite{Adl, BJ} is not modified in $\theta$-expanded gauge theory \cite{BMR}. In other words,
\begin{equation}\label{116}
\mathscr{A} = \partial_{\mu}J^{\mu}_{5} = \frac{1}{16\pi^{2}}\varepsilon_{\mu\nu\lambda\rho}F^{\mu\nu}F^{\lambda\rho}
\end{equation}
still holds. The star-gauge-covariant anomaly is just given by a standard deformation of the above result \cite{AS, GM}:
\begin{equation}\label{117}
\nc{\mathscr{A}} = \nc{\mathrm{D}}_{\mu}\star\nc{J}^{\mu}_{5} = \frac{1}{16\pi^{2}}\varepsilon_{\mu\nu\lambda\rho}\nc{F}^{\mu\nu}\star\nc{F}^{\lambda\rho}.
\end{equation}
The expected map for anomalies, obtained by a lift from the classical result (\ref{115}), follows as
\begin{equation}\label{anomaly4}
\nc{\mathscr{A}} = \mathscr{A}+\theta^{\alpha\beta}\partial_{\alpha}\left(A_{\beta}\mathscr{A}\right)
+\frac{1}{2}\theta^{\alpha\beta}\theta^{\kappa\sigma}\partial_{\alpha}\left[A_{\kappa}F_{\beta\sigma}\mathscr{A}-A_{\beta}\partial_{\sigma}\left(A_{\kappa}\mathscr{A}\right)\right]+O(\theta^{3}).
\end{equation}

Let us digress a bit on this map. The starting point is the classical map (\ref{111}) with the vector current replaced by the axial one. Although current conservation is used to derive the map (\ref{111}), the analysis still remains valid since the axial current is also classically conserved. Also, as discussed earlier, the retention of the term proportional to the divergence of the current would spoil the stability of the gauge transformations, which must hold irrespective of whether the current is vector or axial. From the map (\ref{111}) one is led to the relation (\ref{115}). Now we would like to see whether this classical map persists even at the quantum level, written in the form (\ref{anomaly4}).\footnote{See also the discussion in the last paragraph of Sec.~\ref{sec:map}.} As far as gauge-transformation properties are concerned, it is obviously compatible since the anomalies in the different descriptions transform exactly as the corresponding currents. Corrections, if any, would thus entail only gauge-invariant terms, involving the field tensor $F_{\mu\nu}$. We now prove that the relation (\ref{anomaly4}) is indeed valid for the slowly-varying-field approximation, which was also essential for demonstrating the equivalence of DBI actions \cite{SW}. Later on we shall compute the corrections that appear for arbitrary field configurations. In the slowly-varying-field approximation, since derivatives on $\nc{F}^{\mu\nu}$ can be ignored, the star product in Eq.~(\ref{117}) is dropped. Using the map (\ref{107}), we write this expression as
\begin{equation}\label{120-1}\begin{split}
\nc{\mathscr{A}} &= \frac{1}{16\pi^{2}}\varepsilon_{\mu\nu\lambda\rho}\nc{F}^{\mu\nu}\nc{F}^{\lambda\rho}\\
&= \frac{1}{16\pi^{2}}\varepsilon_{\mu\nu\lambda\rho}\bigg[F^{\mu\nu}F^{\lambda\rho} + \theta^{\alpha\beta}\left\{A_{\beta}\partial_{\alpha}\left(F^{\mu\nu}F^{\lambda\rho}\right)-2F^{\mu\nu}{F^{\lambda}}_{\alpha}{F_{\beta}}^{\rho}\right\}\\
&\qquad\qquad\qquad {}+ \theta^{\alpha\beta}\theta^{\kappa\sigma}\bigg\{\frac{1}{2}A_{\alpha}\partial_{\beta}\left[A_{\kappa}\partial_{\sigma}\left(F^{\mu\nu}F^{\lambda\rho}\right)\right]+\frac{1}{2}A_{\alpha}F_{\beta\kappa}\partial_{\sigma}\left(F^{\mu\nu}F^{\lambda\rho}\right)\\
&\qquad\qquad\qquad\qquad\qquad\quad {}+2A_{\alpha}\partial_{\beta}\left(F^{\mu\nu}{F^{\lambda}}_{\kappa}{F_{\sigma}}^{\rho}\right)+2F^{\mu\nu}{F^{\lambda}}_{\alpha}F_{\beta\kappa}{F_{\sigma}}^{\rho}\\
&\qquad\qquad\qquad\qquad\qquad\quad {}+{F^{\mu}}_{\alpha}{F_{\beta}}^{\nu}{F^{\lambda}}_{\kappa}{F_{\sigma}}^{\rho}\bigg\}+O(\theta^{3})\bigg].\end{split}
\end{equation}
Next, using the identities \cite{Ban}
\begin{gather}
\label{118}
\varepsilon_{\mu\nu\lambda\rho}\theta^{\alpha\beta}\left[F^{\mu\nu}F^{\lambda\rho}F_{\alpha\beta}+4F^{\mu\nu}{F^{\lambda}}_{\alpha}{F_{\beta}}^{\rho}\right] = 0,\\
\label{119}\begin{split}
\varepsilon_{\mu\nu\lambda\rho}\theta^{\alpha\beta}\theta^{\kappa\sigma}&\bigg[{F^{\mu}}_{\alpha}{F_{\beta}}^{\nu}{F^{\lambda}}_{\kappa}{F_{\sigma}}^{\rho}+2F^{\mu\nu}{F^{\lambda}}_{\alpha}F_{\beta\kappa}{F_{\sigma}}^{\rho}\\
&\quad {}+\frac{1}{2}F^{\mu\nu}{F^{\lambda}}_{\kappa}{F_{\sigma}}^{\rho}F_{\alpha\beta}+\frac{1}{4}F^{\mu\nu}F^{\lambda\rho}F_{\alpha\kappa}F_{\sigma\beta}\bigg] = 0,\end{split}
\end{gather}
and the usual Bianchi identity, we can write down
\begin{equation}\label{120}\begin{split}
\nc{\mathscr{A}} &= \frac{1}{16\pi^{2}}\varepsilon_{\mu\nu\lambda\rho}\bigg[F^{\mu\nu}F^{\lambda\rho}+\theta^{\alpha\beta}\partial_{\alpha}\left(A_{\beta}F^{\mu\nu}F^{\lambda\rho}\right)\\
&\qquad\qquad\qquad {}+\frac{1}{2}\theta^{\alpha\beta}\theta^{\kappa\sigma}\partial_{\alpha}\Big\{A_{\kappa}F_{\beta\sigma}F^{\mu\nu}F^{\lambda\rho}-A_{\beta}\partial_{\sigma}\left(A_{\kappa}F^{\mu\nu}F^{\lambda\rho}\right)\Big\}+O(\theta^{3})\bigg].\end{split}
\end{equation}
The identities (\ref{118}) and (\ref{119}) are valid in four dimensions and, in fact, hold not only for just $F^{\mu\nu}$ but for any antisymmetric tensor, in particular, for $\nc{F}^{\mu\nu}$ also. This gives a definite way for obtaining the identity (\ref{119}) starting form (\ref{118}). The identity (\ref{119}) may be obtained from the identity (\ref{118}) by doing the replacement $F^{\mu\nu}\rightarrow\nc{F}^{\mu\nu}$ followed by using the map (\ref{107}) and retaining $O(\theta^{2})$ terms. Alternatively, one can check it by explicitly carrying out all the summations. Now substituting for the anomaly (\ref{116}) on the right-hand side of Eq.~(\ref{120}), we indeed get back our expected anomaly map (\ref{anomaly4}).

It is easy to show that the map (\ref{anomaly4}) is equally valid in two dimensions,\footnote{Contrary to the four-dimensional example, the map holds for arbitrary fields. This is because the anomaly does not involve any (star) product of fields and hence the slowly-varying-field approximation becomes redundant.} in which case,
\begin{equation}\label{an-nc-2d}
\mathscr{A}_{2\mathrm{d}} = \partial_{\mu}J^{\mu}_{5} = \frac{1}{2\pi}\varepsilon_{\mu\nu}F^{\mu\nu},\quad
\nc{\mathscr{A}}_{2\mathrm{d}} = \nc{\mathrm{D}}_{\mu}\star\nc{J}^{\mu}_{5} = \frac{1}{2\pi}\varepsilon_{\mu\nu}\nc{F}^{\mu\nu}.
\end{equation}
It follows from the map (\ref{107}) for the field strength that
\begin{equation}\label{an2-2d}\begin{split}
\nc{\mathscr{A}}_{2\mathrm{d}} &= \frac{1}{2\pi}\varepsilon_{\mu\nu}\nc{F}^{\mu\nu}\\
&= \frac{1}{2\pi}\varepsilon_{\mu\nu}\bigg[F^{\mu\nu}-\theta^{\alpha\beta}\left(A_{\alpha}\partial_{\beta}F^{\mu\nu}+{F^{\mu}}_{\alpha}{F_{\beta}}^{\nu}\right)\\
&\qquad\qquad {}+\frac{1}{2}\theta^{\alpha\beta}\theta^{\kappa\sigma}\left\{A_{\alpha}\partial_{\beta}\left(A_{\kappa}\partial_{\sigma}F^{\mu\nu}\right)+A_{\alpha}F_{\beta\kappa}\partial_{\sigma}F^{\mu\nu}\right.\\
&\qquad\qquad\qquad\qquad\quad \left.{}+2A_{\alpha}\partial_{\beta}\left({F^{\mu}}_{\kappa}{F_{\sigma}}^{\nu}\right)+2{F^{\mu}}_{\alpha}F_{\beta\kappa}{F_{\sigma}}^{\nu}\right\} +O(\theta^{3})\bigg].\end{split}
\end{equation}
In two dimensions, we have the identities
\begin{gather}
\label{id-2d}
\varepsilon_{\mu\nu}\theta^{\alpha\beta}\left(F_{\alpha\beta}F^{\mu\nu}+2{F^{\mu}}_{\alpha}{F_{\beta}}^{\nu}\right) = 0,\\
\label{id2-2d}
\varepsilon_{\mu\nu}\theta^{\alpha\beta}\theta^{\kappa\sigma}\left(F_{\alpha\kappa}F_{\sigma\beta}F^{\mu\nu}+F_{\alpha\beta}{F^{\mu}}_{\kappa}{F_{\sigma}}^{\nu}+4{F^{\mu}}_{\alpha}F_{\beta\kappa}{F_{\sigma}}^{\nu}\right) = 0,
\end{gather}
which are the analogue of the identities (\ref{118}) and (\ref{119}). Likewise, these identities hold for any antisymmetric second-rank tensor, and the second identitiy can be obtained from the first by replacing the usual field strength by the noncommutative field strength and then using the SW map. Using these identities, Eq.~(\ref{an2-2d}) can be rewritten as
\begin{equation}\label{an3-2d}\begin{split}
\nc{\mathscr{A}}_{2\mathrm{d}} = \frac{1}{2\pi}\varepsilon_{\mu\nu}&\bigg[F^{\mu\nu}+\theta^{\alpha\beta}\partial_{\alpha}\left(A_{\beta}F^{\mu\nu}\right)\\
&\quad {}+\frac{1}{2}\theta^{\alpha\beta}\theta^{\kappa\sigma}\partial_{\alpha}\left\{A_{\kappa}F_{\beta\sigma}F^{\mu\nu}-A_{\beta}\partial_{\sigma}\left(A_{\kappa}F^{\mu\nu}\right)\right\}+O(\theta^{3})\bigg],\end{split}
\end{equation}
which, substituting for the usual anomaly on the right-hand side, reproduces the map (\ref{anomaly4}) with $\nc{\mathscr{A}}$ and $\mathscr{A}$ replaced by $\nc{\mathscr{A}}_{2\mathrm{d}}$ and $\mathscr{A}_{2\mathrm{d}}$, respectively. 

For arbitrary fields, the derivative corrections to the map in the four-dimensional case are next computed. Now the noncommutative anomaly takes the form
\begin{equation}\label{120-d}\begin{split}
\nc{\mathscr{A}} &= \frac{1}{16\pi^{2}}\varepsilon_{\mu\nu\lambda\rho}\nc{F}^{\mu\nu}\star\nc{F}^{\lambda\rho}\\
&= \frac{1}{16\pi^{2}}\varepsilon_{\mu\nu\lambda\rho}\bigg[F^{\mu\nu}F^{\lambda\rho}+\theta^{\alpha\beta}\partial_{\alpha}\left(A_{\beta}F^{\mu\nu}F^{\lambda\rho}\right)\\
&\qquad\qquad\qquad {}+\frac{1}{2}\theta^{\alpha\beta}\theta^{\kappa\sigma}\partial_{\alpha}\Big\{A_{\kappa}F_{\beta\sigma}F^{\mu\nu}F^{\lambda\rho}-A_{\beta}\partial_{\sigma}\left(A_{\kappa}F^{\mu\nu}F^{\lambda\rho}\right)\Big\}\bigg]\\
&\quad {}-\frac{1}{128\pi^{2}}\varepsilon_{\mu\nu\lambda\rho}\theta^{\alpha\beta}\theta^{\kappa\sigma}\partial_{\alpha}\partial_{\kappa}F^{\mu\nu}\partial_{\beta}\partial_{\sigma}F^{\lambda\rho}+O(\theta^{3}).\end{split}
\end{equation}
The last term is the new piece added to Eq.~(\ref{120}). Thus, the map (\ref{anomaly4}) gets modified as
\begin{equation}\label{anomaly4-m}\begin{split}
\nc{\mathscr{A}} &= \mathscr{A}+\theta^{\alpha\beta}\partial_{\alpha}\left(A_{\beta}\mathscr{A}\right)
+\frac{1}{2}\theta^{\alpha\beta}\theta^{\kappa\sigma}\partial_{\alpha}\left[A_{\kappa}F_{\beta\sigma}\mathscr{A}-A_{\beta}\partial_{\sigma}\left(A_{\kappa}\mathscr{A}\right)\right]\\
&\quad {}-\frac{1}{128\pi^{2}}\varepsilon_{\mu\nu\lambda\rho}\theta^{\alpha\beta}\theta^{\kappa\sigma}\partial_{\alpha}\left(\partial_{\kappa}F^{\mu\nu}\partial_{\beta}\partial_{\sigma}F^{\lambda\rho}\right)+O(\theta^{3}).\end{split}
\end{equation}
This is reproduced by including a derivative correction to the classical map (\ref{111}) for currents:
\begin{equation}\label{122}\begin{split}
\nc{J}^{\mu}_{5} &= J^{\mu}_{5}-\theta^{\alpha\beta}\left(A_{\alpha}\partial_{\beta}J^{\mu}_{5}-\frac{1}{2}F_{\alpha\beta}J^{\mu}_{5}\right)+\theta^{\mu\alpha}F_{\alpha\beta}J^{\beta}_{5}\\
&\quad {}+\frac{1}{2}\theta^{\alpha\beta}\theta^{\kappa\sigma}\partial_{\alpha}\left(A_{\kappa}F_{\beta\sigma}J^{\mu}_{5}-A_{\beta}A_{\kappa}\partial_{\sigma}J^{\mu}_{5}+\frac{1}{2}A_{\beta}F_{\kappa\sigma}J^{\mu}_{5} \right)-\theta^{\alpha\beta}\theta^{\kappa\mu}\partial_{\alpha}\left(A_{\beta}F_{\kappa\nu}J^{\nu}_{5}\right)\\
&\quad {}+\frac{1}{128\pi^{2}}\varepsilon_{\sigma\nu\lambda\rho}\theta^{\alpha\beta}\theta^{\kappa\mu}\partial_{\alpha}F^{\sigma\nu}\partial_{\kappa}\partial_{\beta}F^{\lambda\rho}+O(\theta^{3}).\end{split}
\end{equation}
The correction term is given at the end. It is straightforward to see the contribution of this derivative term. Since this is an $O(\theta^2)$ term and we are restricting ourselves to the second order itself, taking its noncommutative covariant derivative amounts to just taking its ordinary partial derivative. Then taking into account the antisymmetric nature of $\theta^{\kappa\mu}$ it immediately yields the corresponding term in Eq.~(\ref{anomaly4-m}). We therefore interpret this term as a quantum correction for correctly mapping anomalies for arbitrary fields.

It is to be noted that Eq.~(\ref{anomaly4-m}) can be put in a form so that the $\theta$-dependent terms are all expressed as a total derivative. This implies
\begin{equation}\label{xx}
\int\!\mathrm{d}^{4}x\,\nc{\mathrm{D}}_{\mu}\star\nc{J}^{\mu}_{5} = \int\!\mathrm{d}^{4}x\,\partial_{\mu}J^{\mu}_{5},
\end{equation}
reproducing the familiar equivalence of the integrated anomalies \cite{Ban, AS, GM, BG}.

We shall now give some useful inverse maps. From maps (\ref{110}) and (\ref{111}), the inverse map for the currents follows:
\begin{equation}\label{123}\begin{split}
J^{\mu} &= \nc{J}^{\mu}+\theta^{\alpha\beta}\left(\nc{A}_{\alpha}\partial_{\beta}\nc{J}^{\mu}-\frac{1}{2}\nc{F}_{\alpha\beta}\nc{J}^{\mu}\right)-\theta^{\mu\alpha}\nc{F}_{\alpha\beta}\nc{J}^{\beta}\\
&\quad {}-\frac{1}{2}\theta^{\alpha\beta}\theta^{\kappa\sigma}\left[\nc{A}_{\kappa}\partial_{\beta}\nc{F}_{\sigma\alpha}\nc{J}^{\mu}-\nc{A}_{\alpha}\nc{A}_{\kappa}\partial_{\beta}\partial_{\sigma}\nc{J}^{\mu}-2\nc{A}_{\alpha}\partial_{\beta}\nc{A}_{\kappa}\partial_{\sigma}\nc{J}^{\mu}-\frac{1}{2}\nc{A}_{\kappa}\nc{F}_{\alpha\beta}\partial_{\sigma}\nc{J}^{\mu}\right.\\
&\qquad\qquad\qquad \left. {}+\frac{3}{2}\nc{A}_{\alpha}\partial_{\beta}\left(\nc{F}_{\kappa\sigma}\nc{J}^{\mu}\right)+\frac{1}{2}\nc{F}_{\alpha\kappa}\nc{F}_{\sigma\beta}\nc{J}^{\mu}-\frac{1}{4}\nc{F}_{\alpha\beta}\nc{F}_{\kappa\sigma}\nc{J}^{\mu}\right]\\
&\quad {}-\theta^{\alpha\beta}\theta^{\kappa\mu}\partial_{\alpha}\left(\nc{A}_{\beta}\nc{F}_{\kappa\nu}\nc{J}^{\nu}\right)+O(\theta^{3}).\end{split}
\end{equation}
Taking the ordinary derivative and doing some simplifications yields
\begin{equation}\label{125}
\partial_{\mu}J^{\mu} = \nc{\mathrm{D}}_{\mu}\star\nc{J}^{\mu}-\theta^{\alpha\beta}\partial_{\alpha}\left[\nc{A}_{\beta}\left(\nc{\mathrm{D}}_{\mu}\star\nc{J}^{\mu}\right)\right]+\frac{1}{2}\theta^{\alpha\beta}\theta^{\kappa\sigma}\partial_{\alpha}\partial_{\kappa}\left[\nc{A}_{\beta}\nc{A}_{\sigma}\left(\nc{\mathrm{D}}_{\mu}\star\nc{J}^{\mu}\right)\right]+O(\theta^{3}),
\end{equation}
which may be regarded as the inverse map of (\ref{115}). Indeed, use of this relation reduces the expression on the right-hand side of Eq.~(\ref{115}) to that on its left-hand side which shows the consistency of the results. This also proves that the covariant conservation of $\nc{J}^{\mu}$ implies the ordinary conservation of $J^{\mu}$, as expected.

Likewise, inverting the relation (\ref{107}), we obtain
\begin{equation}\label{127}\begin{split}
F_{\mu\nu} &= \nc{F}_{\mu\nu}+\theta^{\alpha\beta}\left(\nc{A}_{\alpha}\partial_{\beta}\nc{F}_{\mu\nu}+\nc{F}_{\mu\alpha}\nc{F}_{\beta\nu}\right)\\
&\quad {}+\theta^{\alpha\beta}\theta^{\kappa\sigma}\left[\nc{A}_{\alpha}\partial_{\beta}\nc{A}_{\kappa}\partial_{\sigma}\nc{F}_{\mu\nu}+\frac{1}{2}\nc{A}_{\alpha}\nc{A}_{\kappa}\partial_{\beta}\partial_{\sigma}\nc{F}_{\mu\nu}+\nc{A}_{\alpha}\partial_{\beta}\left(\nc{F}_{\mu\kappa}\nc{F}_{\sigma\nu}\right)+\nc{F}_{\mu\alpha}\nc{F}_{\beta\kappa}\nc{F}_{\sigma\nu}\right]\\
&\quad {}+O(\theta^{3}).\end{split}
\end{equation}
If we now write down the usual anomaly as
\begin{equation}\label{126}
\frac{1}{16\pi^{2}}\varepsilon_{\mu\nu\lambda\rho}F^{\mu\nu}F^{\lambda\rho} = \frac{1}{16\pi^{2}}\varepsilon_{\mu\nu\lambda\rho}\left(F^{\mu\nu}\star F^{\lambda\rho}+\frac{1}{8}\theta^{\alpha\beta}\theta^{\kappa\sigma}\partial_{\alpha}\partial_{\kappa}F^{\mu\nu}\partial_{\beta}\partial_{\sigma}F^{\lambda\rho}+O(\theta^{3})\right),
\end{equation}
and use Eq.~(\ref{127}) on the right-hand side, we get
\begin{equation}\label{128}\begin{split}
\frac{1}{16\pi^{2}}\varepsilon_{\mu\nu\lambda\rho}F^{\mu\nu}F^{\lambda\rho} &= \frac{1}{16\pi^{2}}\varepsilon_{\mu\nu\lambda\rho}\bigg[\nc{F}^{\mu\nu}\star\nc{F}^{\lambda\rho}-\theta^{\alpha\beta}\partial_{\alpha}\left\{\nc{A}_{\beta}\left(\nc{F}^{\mu\nu}\star\nc{F}^{\lambda\rho}\right)\right\}\\
&\qquad\qquad\qquad {}+\frac{1}{2}\theta^{\alpha\beta}\theta^{\kappa\sigma}\partial_{\alpha}\partial_{\kappa}\Big\{\nc{A}_{\beta}\nc{A}_{\sigma}\left(\nc{F}^{\mu\nu}\star\nc{F}^{\lambda\rho}\right)\Big\}\bigg]\\
&\quad {}+\frac{1}{128\pi^{2}}\varepsilon_{\mu\nu\lambda\rho}\theta^{\alpha\beta}\theta^{\kappa\sigma}\partial_{\alpha}\partial_{\kappa}\nc{F}^{\mu\nu}\partial_{\beta}\partial_{\sigma}\nc{F}^{\lambda\rho}+O(\theta^{3}),\end{split}
\end{equation}
where we have used the identities (\ref{118}) and (\ref{119}) with the replacement  $F^{\mu\nu}\rightarrow\nc{F}^{\mu\nu}$. Thus we have the map for the anomalies:
\begin{equation}\label{129}\begin{split}
\partial_{\mu}J^{\mu}_{5} &= \nc{\mathrm{D}}_{\mu}\star\nc{J}^{\mu}_{5}-\theta^{\alpha\beta}\partial_{\alpha}\left[\nc{A}_{\beta}\left(\nc{\mathrm{D}}_{\mu}\star\nc{J}^{\mu}_{5}\right)\right]+\frac{1}{2}\theta^{\alpha\beta}\theta^{\kappa\sigma}\partial_{\alpha}\partial_{\kappa}\left[\nc{A}_{\beta}\nc{A}_{\sigma}\left(\nc{\mathrm{D}}_{\mu}\star\nc{J}^{\mu}_{5}\right)\right]\\
&\quad {}+\frac{1}{128\pi^{2}}\varepsilon_{\mu\nu\lambda\rho}\theta^{\alpha\beta}\theta^{\kappa\sigma}\partial_{\alpha}\partial_{\kappa}\nc{F}^{\mu\nu}\partial_{\beta}\partial_{\sigma}\nc{F}^{\lambda\rho}+O(\theta^{3}).\end{split}
\end{equation}
In the slowly-varying-field approximation, the last term drops out. Then it mimics the usual map (\ref{125}). Again, as before, it is possible to find the correction term for arbitrary fields and write down the map for anomalous current as
\begin{equation}\label{130}\begin{split}
J^{\mu}_{5} &= \nc{J}^{\mu}_{5}+\theta^{\alpha\beta}\left(\nc{A}_{\alpha}\partial_{\beta}\nc{J}^{\mu}_{5}-\frac{1}{2}\nc{F}_{\alpha\beta}\nc{J}^{\mu}_{5}\right)-\theta^{\mu\alpha}\nc{F}_{\alpha\beta}\nc{J}^{\beta}_{5}\\
&\quad {}-\frac{1}{2}\theta^{\alpha\beta}\theta^{\kappa\sigma}\left[\nc{A}_{\kappa}\partial_{\beta}\nc{F}_{\sigma\alpha}\nc{J}^{\mu}_{5}-\nc{A}_{\alpha}\nc{A}_{\kappa}\partial_{\beta}\partial_{\sigma}\nc{J}^{\mu}_{5}-2\nc{A}_{\alpha}\partial_{\beta}\nc{A}_{\kappa}\partial_{\sigma}\nc{J}^{\mu}_{5}-\frac{1}{2}\nc{A}_{\kappa}\nc{F}_{\alpha\beta}\partial_{\sigma}\nc{J}^{\mu}_{5}\right.\\
&\qquad\qquad\qquad \left. {}+\frac{3}{2}\nc{A}_{\alpha}\partial_{\beta}\left(\nc{F}_{\kappa\sigma}\nc{J}^{\mu}_{5}\right)+\frac{1}{2}\nc{F}_{\alpha\kappa}\nc{F}_{\sigma\beta}\nc{J}^{\mu}_{5}-\frac{1}{4}\nc{F}_{\alpha\beta}\nc{F}_{\kappa\sigma}\nc{J}^{\mu}_{5}\right]\\
&\quad {}-\theta^{\alpha\beta}\theta^{\kappa\mu}\partial_{\alpha}\left(\nc{A}_{\beta}\nc{F}_{\kappa\nu}\nc{J}^{\nu}_{5}\right)-\frac{1}{128\pi^{2}}\varepsilon_{\sigma\nu\lambda\rho}\theta^{\alpha\beta}\theta^{\kappa\mu}\partial_{\alpha}\nc{F}^{\sigma\nu}\partial_{\kappa}\partial_{\beta}\nc{F}^{\lambda\rho}+O(\theta^{3}),\end{split}
\end{equation}
which reproduces Eq.~(\ref{129}) correctly. Substituting this map, the expression on the right-hand side of Eq.~(\ref{122}) reduces to that on its left-hand side, which shows the consistency of the results.

We conclude this section by providing a mapping between modified chiral currents which are anomaly-free but no longer gauge-invariant. In the ordinary (commutative) theory, such a modified chiral current may be defined as
\begin{equation}\label{Jcal}
\mathcal{J}^{\mu} = J^{\mu}_{5}-\frac{1}{8\pi^{2}}\varepsilon^{\mu\nu\lambda\rho}A_{\nu}F_{\lambda\rho}.
\end{equation}
By construction, this is anomaly-free ($\partial_{\mu}\mathcal{J}^{\mu}=0$) but no longer gauge-invariant. It is possible to do a similar thing for the noncommutative theory. We rewrite Eq.~(\ref{122}) by replacing $J^{\mu}_{5}$ in favour of $\mathcal{J}^{\mu}$. The terms independent of $\mathcal{J}^{\mu}$, including the quantum correction, are then moved to the other side and a new current is defined as
\begin{equation}\label{Jcalhat}
\nc{\mathcal{J}}^{\mu} = \nc{J}^{\mu}_{5}+\nc{X}^{\mu}(\nc{A}),
\end{equation}
where all $A_{\mu}$-dependent terms lumped in $\nc{X}^{\mu}$ have been expressed in terms of the noncommutative variables using the SW map. Thus we have
\begin{equation}\label{111cal}\begin{split}
\nc{\mathcal{J}}^{\mu} &= \mathcal{J}^{\mu}-\theta^{\alpha\beta}\left(A_{\alpha}\partial_{\beta}\mathcal{J}^{\mu}-\frac{1}{2}F_{\alpha\beta}\mathcal{J}^{\mu}\right)+\theta^{\mu\alpha}F_{\alpha\beta}\mathcal{J}^{\beta}\\
&\quad {}+\frac{1}{2}\theta^{\alpha\beta}\theta^{\kappa\sigma}\partial_{\alpha}\left(A_{\kappa}F_{\beta\sigma}\mathcal{J}^{\mu}-A_{\beta}A_{\kappa}\partial_{\sigma}\mathcal{J}^{\mu}+\frac{1}{2}A_{\beta}F_{\kappa\sigma}\mathcal{J}^{\mu}\right)
-\theta^{\alpha\beta}\theta^{\kappa\mu}\partial_{\alpha}\left(A_{\beta}F_{\kappa\nu}\mathcal{J}^{\nu}\right)\\
&\quad {}+O(\theta^{3}).\end{split}
\end{equation}
Since the above equation is structurally identical to Eq.~(\ref{111}), a relation akin to (\ref{115}) follows:
\begin{equation}\label{115cal}\begin{split}
\nc{\mathrm{D}}_{\mu}\star\nc{\mathcal{J}}^{\mu} &= \partial_{\mu}\mathcal{J}^{\mu}+\theta^{\alpha\beta}\partial_{\alpha}\left(A_{\beta}\partial_{\mu}\mathcal{J}^{\mu}\right)\\
&\quad {}+\frac{1}{2}\theta^{\alpha\beta}\theta^{\kappa\sigma}\partial_{\alpha}\left[A_{\kappa}F_{\beta\sigma}\partial_{\mu}\mathcal{J}^{\mu}-A_{\beta}\partial_{\sigma}\left(A_{\kappa}\partial_{\mu}\mathcal{J}^{\mu}\right)\right]+O(\theta^{3}),\end{split}
\end{equation}
which shows that $\partial_{\mu}\mathcal{J}^{\mu}=0$ implies $\nc{\mathrm{D}}_{\mu}\star\nc{\mathcal{J}}^{\mu}=0$. We are thus successful in constructing  an anomaly-free current which however does not transform (star-) covariantly. It is the $\nc{X}^{\mu}$, appearing in Eq.~(\ref{Jcalhat}), which spoils the covariance of $\nc{\mathcal{J}}^{\mu}$.

%%%%%%%%%%%%%%%%%%%%%%%%%%%%%%%%%%%%%%%%%%%%%%%%%%%%%%%%%%%%%%%%%%%%%%%%%%%%%%%

\section{\label{sec:higher}Higher-order computations}

Results in the previous section were valid up to $O(\theta^{2})$. A natural question that arises is the validity of these results for further higher-order corrections. Here we face a problem. The point is that although the map (\ref{109}) for sources is given in a closed form, its explicit structure is dictated by the map involving the potentials. Thus one has to first construct the latter map before proceeding. All these features make higher- (than $O(\theta^{2})$) order computations very formidable, if not practically impossible. An alternate approach is suggested, which is explicitly demonstrated by considering $O(\theta^{3})$ calculations.

Consider first the two-dimensional example. The star-gauge-covariant anomaly, after an application of the SW map, is given by
\begin{equation}\label{F-3.m1}
\nc{\mathscr{A}}_{2\mathrm{d}} = \frac{1}{2\pi}\varepsilon_{\mu\nu}\nc{F}^{\mu\nu} = {\mathscr{A}}_{2\mathrm{d}}^{(0)}+{\mathscr{A}}_{2\mathrm{d}}^{(1)}+{\mathscr{A}}_{2\mathrm{d}}^{(2)}+{\mathscr{A}}_{2\mathrm{d}}^{(3)}+O(\theta^{4}),
\end{equation}
with ${\mathscr{A}}_{2\mathrm{d}}^{(0)}$, ${\mathscr{A}}_{2\mathrm{d}}^{(1)}$ and ${\mathscr{A}}_{2\mathrm{d}}^{(2)}$ respectively being the zeroth-, first- and second-order (in $\theta$) parts already appearing on the right-hand side of Eq.~(\ref{an3-2d}), and
\begin{equation}\label{F-3.m}\begin{split}
{\mathscr{A}}_{2\mathrm{d}}^{(3)} &= - \frac{1}{12\pi}\varepsilon_{\mu\nu}\theta^{\alpha\beta}\theta^{\kappa\sigma}\theta^{\tau\xi}\\
&\qquad \times\left[A_{\alpha}\partial_{\beta}\left\{A_{\kappa}\partial_{\sigma}\left(A_{\tau}\partial_{\xi}F^{\mu\nu}+3{F^{\mu}}_{\tau}{F_{\xi}}^{\nu}\right)+2F_{\sigma\tau}\left(A_{\kappa}\partial_{\xi}F^{\mu\nu}+3{F^{\mu}}_{\kappa}{F_{\xi}}^{\nu}\right)\right\}\right.\\
&\qquad\quad \left.{}+A_{\alpha}F_{\beta\kappa}\partial_{\sigma}\left(A_{\tau}\partial_{\xi}F^{\mu\nu}+3{F^{\mu}}_{\tau}{F_{\xi}}^{\nu}\right)+2F_{\beta\kappa}F_{\sigma\tau}\left(A_{\alpha}\partial_{\xi}F^{\mu\nu}+3{F^{\mu}}_{\alpha}{F_{\xi}}^{\nu}\right)\right],\end{split}
\end{equation}
where the $O(\theta^{3})$ contribution to the map (\ref{107}) has been taken from Ref.~\cite{Fid}.

Now our objective is to rewrite the $O(\theta^{3})$ contribution in a form akin to $O(\theta)$ and $O(\theta^{2})$ terms; namely, to recast it as something proportional to the commutative anomaly ($\varepsilon_{\mu\nu}F^{\mu\nu}$), and also as a total derivative. Expressing it as a total derivative is necessary to preserve the equality of the integrated anomalies ($\int\!\mathrm{d}^{2}x\,\varepsilon_{\mu\nu}\nc{F}^{\mu\nu} = \int\!\mathrm{d}^{2}x\,\varepsilon_{\mu\nu}F^{\mu\nu}$) \cite{BY, BLY, Ban, AS, GM}.

The $O(\theta^{3})$ contribution may be expressed as
\begin{equation}\label{an-o3.2d}\begin{split}
{\mathscr{A}}_{2\mathrm{d}}^{(3)} &= -\frac{1}{12\pi}\varepsilon_{\mu\nu}\theta^{\alpha\beta}\theta^{\kappa\sigma}\theta^{\tau\xi}\left[A_{\alpha}\partial_{\beta}\left\{A_{\kappa}\partial_{\sigma}\left(A_{\tau}\partial_{\xi}F^{\mu\nu}-\frac{3}{2}F_{\tau\xi}F^{\mu\nu}\right)+2A_{\kappa}F_{\sigma\tau}\partial_{\xi}F^{\mu\nu}\right.\right.\\
&\qquad\qquad\qquad\qquad\qquad\qquad\qquad \left.{}+\frac{3}{4}\left(F_{\kappa\sigma}F_{\tau\xi}-2F_{\kappa\tau}F_{\xi\sigma}\right)F^{\mu\nu}\right\}\\
&\qquad\qquad\qquad\qquad\qquad {}+A_{\alpha}F_{\beta\kappa}\left\{\partial_{\sigma}\left(A_{\tau}\partial_{\xi}F^{\mu\nu}-\frac{3}{2}F_{\tau\xi}F^{\mu\nu}\right)+2F_{\sigma\tau}\partial_{\xi}F^{\mu\nu}\right\}\\
&\qquad\qquad\qquad\qquad\qquad \left.{}-\left(F_{\alpha\tau}F_{\xi\kappa}F_{\sigma\beta}+\frac{1}{8}F_{\alpha\beta}F_{\kappa\sigma}F_{\tau\xi}-\frac{3}{4}F_{\alpha\beta}F_{\kappa\tau}F_{\xi\sigma}\right)F^{\mu\nu}\right],\end{split}
\end{equation}
where, in addition to the identities (\ref{id-2d}) and (\ref{id2-2d}), we have also used
\begin{equation}\label{id3-2d}\begin{split}
\varepsilon_{\mu\nu}\theta^{\alpha\beta}\theta^{\kappa\sigma}\theta^{\tau\xi}&\left(F^{\mu\nu}F_{\alpha\tau}F_{\xi\kappa}F_{\sigma\beta}+{F^{\mu}}_{\kappa}{F_{\sigma}}^{\nu}F_{\alpha\tau}F_{\xi\beta}\right.\\
&\left.{}+{F^{\mu}}_{\kappa}F_{\sigma\tau}{F_{\xi}}^{\nu}F_{\alpha\beta}+6{F^{\mu}}_{\alpha}F_{\beta\kappa}F_{\sigma\tau}{F_{\xi}}^{\nu}\right) =0,\end{split}
\end{equation}
which follows from the identity (\ref{id2-2d}) by doing the replacement $F^{\mu\nu}\rightarrow\nc{F}^{\mu\nu}$ followed by exploiting the SW map and retaining $O(\theta^{3})$ terms. We notice that each term on the right-hand side of Eq.~(\ref{an-o3.2d}) contains the usual anomaly, as desired. After some algebra, the right-hand side of Eq.~(\ref{an-o3.2d}) can be written as a total divergence, which gives us the final improved version of the map (\ref{an3-2d}) as
\begin{equation}\label{an5-2d}\begin{split}
\nc{\mathscr{A}}_{2\mathrm{d}} &= \frac{1}{2\pi}\varepsilon_{\mu\nu}\bigg[F^{\mu\nu}+\theta^{\alpha\beta}\partial_{\alpha}\left(A_{\beta}F^{\mu\nu}\right)+\frac{1}{2}\theta^{\alpha\beta}\theta^{\kappa\sigma}\partial_{\alpha}\left\{A_{\kappa}F_{\beta\sigma}F^{\mu\nu}-A_{\beta}\partial_{\sigma}\left(A_{\kappa}F^{\mu\nu}\right)\right\}\\
&\qquad\qquad\quad {}+\frac{1}{6}\theta^{\alpha\beta}\theta^{\kappa\sigma}\theta^{\tau\xi}\partial_{\alpha}\bigg\{F^{\mu\nu}\bigg(2A_{\tau}F_{\xi\kappa}F_{\sigma\beta}-2A_{\beta}A_{\kappa}\partial_{\sigma}F_{\tau\xi}-\frac{3}{2}A_{\beta}F_{\kappa\tau}F_{\xi\sigma}\\
&\qquad\qquad\qquad\qquad\qquad\qquad\qquad\quad {}+\frac{1}{4}A_{\beta}F_{\kappa\sigma}F_{\tau\xi}-A_{\beta}\partial_{\sigma}\left(A_{\tau}F_{\xi\kappa}\right)-\frac{1}{2}A_{\kappa}F_{\sigma\beta}F_{\tau\xi}\bigg)\\
&\qquad\qquad\qquad\qquad\qquad\qquad\quad {}+\partial_{\xi}F^{\mu\nu}\left[A_{\beta}A_{\kappa}\left(\partial_{\sigma}A_{\tau}+2F_{\sigma\tau}\right)-A_{\tau}\left(A_{\kappa}F_{\beta\sigma}+A_{\beta}F_{\kappa\sigma}\right)\right]\\
&\qquad\qquad\qquad\qquad\qquad\qquad\quad {}+A_{\beta}A_{\kappa}A_{\tau}\partial_{\sigma}\partial_{\xi}F^{\mu\nu}\bigg\} +O(\theta^{4})\bigg].\end{split}
\end{equation}
Thus, in two dimensions, the noncommutative anomaly can be written in terms of the usual anomaly at $O(\theta^{3})$ also:
\begin{equation}\label{an51-2d}\begin{split}
\nc{\mathscr{A}}_{2\mathrm{d}} &= \mathscr{A}_{2\mathrm{d}}+\theta^{\alpha\beta}\partial_{\alpha}\left(A_{\beta}\mathscr{A}_{2\mathrm{d}}\right)+\frac{1}{2}\theta^{\alpha\beta}\theta^{\kappa\sigma}\partial_{\alpha}\left\{A_{\kappa}F_{\beta\sigma}\mathscr{A}_{2\mathrm{d}}-A_{\beta}\partial_{\sigma}\left(A_{\kappa}\mathscr{A}_{2\mathrm{d}}\right)\right\}\\
&\quad {}+\frac{1}{6}\theta^{\alpha\beta}\theta^{\kappa\sigma}\theta^{\tau\xi}\partial_{\alpha}\bigg[\mathscr{A}_{2\mathrm{d}}\bigg(2A_{\tau}F_{\xi\kappa}F_{\sigma\beta}-2A_{\beta}A_{\kappa}\partial_{\sigma}F_{\tau\xi}-\frac{3}{2}A_{\beta}F_{\kappa\tau}F_{\xi\sigma}\\
&\qquad\qquad\qquad\qquad\qquad\quad {}+\frac{1}{4}A_{\beta}F_{\kappa\sigma}F_{\tau\xi}-A_{\beta}\partial_{\sigma}\left(A_{\tau}F_{\xi\kappa}\right)-\frac{1}{2}A_{\kappa}F_{\sigma\beta}F_{\tau\xi}\bigg)\\
&\qquad\qquad\qquad\qquad\quad {}+\partial_{\xi}\mathscr{A}_{2\mathrm{d}}\left\{A_{\beta}A_{\kappa}\left(\partial_{\sigma}A_{\tau}+2F_{\sigma\tau}\right)-A_{\tau}\left(A_{\kappa}F_{\beta\sigma}+A_{\beta}F_{\kappa\sigma}\right)\right\}\\
&\qquad\qquad\qquad\qquad\quad {}+A_{\beta}A_{\kappa}A_{\tau}\partial_{\sigma}\partial_{\xi}\mathscr{A}_{2\mathrm{d}}\bigg] +O(\theta^{4}).\end{split}
\end{equation}

If the anomalies in four dimensions also satisfy the above map, then clearly we have a general result, valid up to $O(\theta^{3})$. Now it will be shown that, in the slowly-varying-field approximation, such a relation indeed holds. We have
\begin{equation}\label{an5-4d}\begin{split}
\lefteqn{\frac{1}{16\pi^{2}}\varepsilon_{\mu\nu\lambda\rho}\nc{F}^{\mu\nu}\nc{F}^{\lambda\rho}}\\
&= \frac{1}{16\pi^{2}}\varepsilon_{\mu\nu\lambda\rho}\bigg[F^{\mu\nu}F^{\lambda\rho}+\theta^{\alpha\beta}\partial_{\alpha}\left(A_{\beta}F^{\mu\nu}F^{\lambda\rho}\right)\\
&\qquad\qquad\qquad {}+\frac{1}{2}\theta^{\alpha\beta}\theta^{\kappa\sigma}\partial_{\alpha}\left\{A_{\kappa}F_{\beta\sigma}F^{\mu\nu}F^{\lambda\rho}-A_{\beta}\partial_{\sigma}\left(A_{\kappa}F^{\mu\nu}F^{\lambda\rho}\right)\right\}\\
&\qquad\qquad\qquad {}+\frac{1}{6}\theta^{\alpha\beta}\theta^{\kappa\sigma}\theta^{\tau\xi}\partial_{\alpha}\bigg\{F^{\mu\nu}F^{\lambda\rho}\bigg(2A_{\tau}F_{\xi\kappa}F_{\sigma\beta}-2A_{\beta}A_{\kappa}\partial_{\sigma}F_{\tau\xi}-\frac{3}{2}A_{\beta}F_{\kappa\tau}F_{\xi\sigma}\\
&\qquad\qquad\qquad\qquad\qquad\qquad\qquad\qquad\qquad {}+\frac{1}{4}A_{\beta}F_{\kappa\sigma}F_{\tau\xi}-A_{\beta}\partial_{\sigma}\left(A_{\tau}F_{\xi\kappa}\right)-\frac{1}{2}A_{\kappa}F_{\sigma\beta}F_{\tau\xi}\bigg)\\
&\qquad\qquad\qquad\qquad\qquad\qquad\qquad {}+\partial_{\xi}\left(F^{\mu\nu}F^{\lambda\rho}\right)\left[A_{\beta}A_{\kappa}\left(\partial_{\sigma}A_{\tau}+2F_{\sigma\tau}\right)\right.\\
&\qquad\qquad\qquad\qquad\qquad\qquad\qquad\qquad\qquad\qquad\quad \left.{}
-A_{\tau}\left(A_{\kappa}F_{\beta\sigma}+A_{\beta}F_{\kappa\sigma}\right)\right]\\
&\qquad\qquad\qquad\qquad\qquad\qquad\qquad {}+A_{\beta}A_{\kappa}A_{\tau}\partial_{\sigma}\partial_{\xi}\left(F^{\mu\nu}F^{\lambda\rho}\right)\bigg\} +O(\theta^{4})\bigg].\end{split}
\end{equation}\label{id3-4d}
In obtaining this equation, it is necessary to use the identities (\ref{118}) and (\ref{119}), and a new one (given below), which follows from the identity (\ref{119}) by doing the replacement $F^{\mu\nu}\rightarrow\nc{F}^{\mu\nu}$ followed by using the SW map and retaining $O(\theta^{3})$ terms:
\begin{equation}\begin{split}
\varepsilon_{\mu\nu\lambda\rho}\theta^{\alpha\beta}\theta^{\kappa\sigma}\theta^{\tau\xi}&\bigg(6{F^{\mu}}_{\alpha}{F_{\beta}}^{\nu}{F^{\lambda}}_{\kappa}F_{\sigma\tau}{F_{\xi}}^{\rho}+6F^{\mu\nu}{F^{\lambda}}_{\alpha}F_{\beta\kappa}F_{\sigma\tau}{F_{\xi}}^{\rho}\\
&\quad {}+F^{\mu\nu}{F^{\lambda}}_{\kappa}{F_{\sigma}}^{\rho}F_{\alpha\tau}F_{\xi\beta}+F^{\mu\nu}{F^{\lambda}}_{\kappa}F_{\sigma\tau}{F_{\xi}}^{\rho}F_{\alpha\beta}\\
&\quad {}+\frac{1}{2}{F^{\mu}}_{\tau}{F_{\xi}}^{\nu}{F^{\lambda}}_{\kappa}{F_{\sigma}}^{\rho}F_{\alpha\beta}+\frac{1}{2}F^{\mu\nu}F^{\lambda\rho}F_{\alpha\tau}F_{\xi\kappa}F_{\sigma\beta}\bigg) = 0.\end{split}
\end{equation}
Obviously, Eq.~(\ref{an5-4d}) reproduces the map (\ref{an51-2d}), with $\nc{\mathscr{A}}_{2\mathrm{d}}$ and $\mathscr{A}_{2\mathrm{d}}$ replaced by the corresponding expressions in four dimensions. This proves our claim.

Starting from the results in two dimensions, it is thus feasible to infer the general structure valid in higher dimensions. This is an outcome of the topological properties of anomalies. Proceeding in this fashion, the map for the anomalies can be extended to higher orders.

%%%%%%%%%%%%%%%%%%%%%%%%%%%%%%%%%%%%%%%%%%%%%%%%%%%%%%%%%%%%%%%%%%%%%%%%%%%%%%%

\section{Discussions\label{sec:conc}}

We have provided a SW-type map relating the sources in the noncommutative and commutative descriptions. In the non-Abelian theory, the classical maps for the currents and their covariant divergences were given up to $O(\theta)$. For investigating quantum aspects of the mapping, we applied it to the divergence anomalies for the Abelian theory in the two descriptions. For the slowly-varying-field approximation, the anomalies indeed got identified. Thus the classical map correctly accounted for the quantum effects inherent in the calculations of the anomalies. The results were checked up to $O(\theta^{2})$. We also provided an indirect method of extending the calculations and found an agreement up to $O(\theta^{3})$. Our analysis strongly suggests that the classical mapping would hold for all orders in $\theta$, albeit in the slowly-varying-field approximation. Our findings may also be compared with Refs.~\cite{KOO, Kam} where the classical equivalence of the Chern--Simons theories in different descriptions was found to persist even in the quantum case.

For arbitrary field configurations, derivative corrections to the classical source map were explicitly computed up to $O(\theta^{2})$. Indeed, it is known that if one has to go beyond the slowly-varying-field approximation, derivative corrections are essential. For instance, DBI actions with derivative corrections have been discussed \cite{Wyll-npb, Wyll-jhep, DMS}. A possible extension of this analysis would be to study the mapping of conformal (trace) anomalies.

To put our results in a proper perspective, let us recall that the SW maps are classical maps. A priori, therefore, it was not clear whether they had any role in the mapping of anomalies which are essentially of quantum origin. The first hint that such a possibility might exist came from Eq.~(\ref{115}) (or Eq.~(\ref{anomaly4})) where the covariant derivative of the noncommutative covariant current was expressed in terms of the ordinary derivative of the commutative current. Indeed, to put the map in this form was quite non-trivial. Classically, such a map was trivially consistent, since both the covariant divergence in the noncommutative description and the ordinary divergence in the usual (commutative) picture vanish. The remarkable feature, however, was that such a map remained valid even for the quantum case in the slowly-varying-field approximation which was checked explicitly by inserting the familiar anomalies\footnote{The planar anomaly for the noncommutative description and the ABJ anomaly for the commutative case.} in the different descriptions. Incidentally, the slowly-varying-field approximation is quite significant in discussions of the SW maps. For instance, it was in this approximation that the equivalence of the DBI actions in the noncommutative and the commutative pictures was established \cite{SW} through the use of SW maps.

Our analysis has certain implications for the mapping among the effective actions (for chiral theories) obtained by integrating out the matter degrees of freedom. The point is that the anomalies are the gauge-variations of the effective actions and if the anomalies get mapped then one expects that, modulo local counterterms, the effective actions might get identified, i.e., it suggests that
\begin{equation}\label{zzz}
\nc{W}\left(\nc{A}(A)\right) \equiv W(A) + \mbox{local counterterms},
\end{equation}
where $W$ and $\nc{W}$ denote the effective actions in the commutative and noncommutative formulations, respectively. Taking the gauge-variations (with parameters $\lambda$ and $\nc{\lambda}$), yields
\begin{equation}\label{xxx2}
\int\!\mathrm{d}^{4}x\,\left(\nc{\mathrm{D}}_{\mu}\star\nc{J}^{\mu}_{5}\right)\star\nc{\lambda} = \int\!\mathrm{d}^{4}x\,\left(\partial_{\mu}J^{\mu}_{5}\right)\lambda+\int\!\mathrm{d}^{4}x\,\left(\partial_{\mu}\Lambda^{\mu}\right)\lambda,
\end{equation}
where
\begin{equation}\label{yyy}
\nc{J}^{\mu}_{5} = \frac{\delta\nc{W}}{\delta\nc{A}_{\mu}}, \quad J^{\mu}_{5} = \frac{\delta W}{\delta A_{\mu}}
\end{equation}
and $\Lambda^{\mu}$ accounts for the ambiguity (local counterterms) in obtaining the effective actions. Now Eq.~(\ref{120-d}) expresses the noncommutative anomaly in terms of the commutative variables. Using that result and the SW map (\ref{102}) for the gauge parameter $\nc{\lambda}$ simplifies the left-hand side of Eq.~(\ref{xxx2}):
\begin{equation}\label{qqq}\begin{split}
\int\!\mathrm{d}^{4}x\,\left(\nc{\mathrm{D}}_{\mu}\star\nc{J}^{\mu}_{5}\right)\star\nc{\lambda} &= \int\!\mathrm{d}^{4}x\,\left(\nc{\mathrm{D}}_{\mu}\star\nc{J}^{\mu}_{5}\right)\nc{\lambda} = \frac{1}{16\pi^{2}}\varepsilon_{\mu\nu\lambda\rho}\int\!\mathrm{d}^{4}x\,\left(\nc{F}^{\mu\nu}\star\nc{F}^{\lambda\rho}\right)\nc{\lambda}\\
&= \frac{1}{16\pi^{2}}\varepsilon_{\mu\nu\lambda\rho}\int\!\mathrm{d}^{4}x\,\left(F^{\mu\nu}F^{\lambda\rho}\right)\lambda+\int\!\mathrm{d}^{4}x\,\left(\partial_{\alpha}\Lambda^{\alpha}\right)\lambda,\end{split}
\end{equation}
where Eq.~(\ref{120-d}) and the  map (\ref{102}) have been used in the last step, and
\begin{equation}\label{Lambda}\begin{split}
\Lambda^{\alpha} &= \frac{1}{16\pi^{2}}\varepsilon_{\mu\nu\lambda\rho}\left[\frac{1}{2}\theta^{\alpha\beta}A_{\beta}F^{\mu\nu}F^{\lambda\rho}+\theta^{\alpha\beta}\theta^{\kappa\sigma}\left(\frac{1}{3}A_{\kappa}F_{\beta\sigma}F^{\mu\nu}F^{\lambda\rho}+\frac{1}{6}A_{\beta}\partial_{\kappa}\left(A_{\sigma}F^{\mu\nu}F^{\lambda\rho}\right)\right.\right.\\
&\qquad\qquad\qquad\qquad\qquad\qquad\qquad\qquad\qquad \left.\left.{}-\frac{1}{8}\partial_{\kappa}F^{\mu\nu}\partial_{\beta}\partial_{\sigma}F^{\lambda\rho}\right)\right],\end{split}
\end{equation}
thereby proving Eq.~(\ref{xxx2}) and establishing the claim (\ref{zzz}).

We further stress, to avoid any confusion, that the relation (\ref{zzz}) was not assumed, either explicitly or implicitly, in our calculations.\footnote{Indeed, as already stated, there cannot be any a priory basis for such an assumption since the classical SW map need not be valid for mapping effective actions that take into account loop effects.} Rather, as shown here, our analysis suggested such a relation. Its explicit verification confirms the consistency of our approach. It should be mentioned that the map among anomalies (\ref{anomaly4}) follows from the map (\ref{111}) for currents through a series of algebraic manipulations. This does not depend on the interpretation of the anomaly as gauge-variation of an effective action. If one sticks to this interpretation and furthermore \emph{assumes} the relation (\ref{zzz}), then it might be possible to get a relation (like Eq.~(\ref{xxx2})) involving the integrated version of the products of anomalies and gauge parameters. Our formulation always led to maps involving unintegrated anomalies or currents, which are more fundamental.

We also note that the map (\ref{anomaly4}) for the unintegrated anomalies, which follows from the basic map (\ref{111}) among the currents, was only valid in the slowly-varying-field approximation. The suggested map (\ref{zzz}) among the effective actions, on the other hand, led to the map (\ref{xxx2}), involving the integrated anomalies and the gauge parameters, that was valid in general. For the pure integrated anomalies we have the familiar map (\ref{xx}) that has been discussed extensively in the literature \cite{Ban, AS, GM, BG}.

%%%%%%%%%%%%%%%%%%%%%%%%%%%%%%%%%%%%%%%%%%%%%%%%%%%%%%%%%%%%%%%%%%%%%%%%%%%%%%%

\section*{Acknowledgments}

K.~K.~thanks the Council of Scientific and Industrial Research (CSIR), Government of India for financial support.

%%%%%%%%%%%%%%%%%%%%%%%%%%%%%%%%%%%%%%%%%%%%%%%%%%%%%%%%%%%%%%%%%%%%%%%%%%%%%%%

%%%%%%%%%%%%%%%%%%%%%%%%%%%%%%%%%%%%%%%%%%%%%%%%%%%%%%%%%%%%%%%%%%%%%%%%%%%%%%%

\end{document}